\documentclass[letterpaper,english,nofootinbib]{revtex4}
\usepackage{newcent}

\usepackage[T1]{fontenc}
\usepackage[latin9]{inputenc}
\usepackage{lmodern}
\usepackage[latin9]{inputenc}
\usepackage[letterpaper]{geometry}
\usepackage{amsmath}
\usepackage{bbm}
\usepackage{graphicx}
\usepackage{color}
\usepackage{babel}
\usepackage{amssymb}

\begin{document}

\title{Longitudinal and transverse relativity of spacetime locality in Planck-scale-deformed phase spaces}

\author{Niccol\'o LORET\footnote{niccolo.loret@roma1.infn.it}}
\author{Leonardo~BARCAROLI }
\author{Giacomo~ROSATI\footnote{giacomo.rosati@roma1.infn.it}}

\affiliation{Dipartimento di Fisica, Universit\`a di Roma ``La Sapienza", P.le A. Moro 2, 00185 Roma, EU\\
and INFN, Sez.~Roma1, P.le A. Moro 2, 00185 Roma, EU}

\begin{abstract}
\noindent
We summarize some of the results we obtained in arXiv:1006.2126 (Physical Review Letters 106, 071301), arXiv:1102.4637 (Physics Letters B 700, 150-156)
and in arXiv:1107.3334, giving complementary characterizations of the relativity of spacetime
locality that affects certain Planck-scale-deformed phase-space descriptions of free particles.
\end{abstract}
\maketitle

\nopagebreak

\section{Preliminaries}

These notes are based on our works in Refs.~\cite{bob,kbob,transverse}. We here plan to give
a quick overall picture of these results that we derived concerning manifestations
of relative locality for theories of free particles  
 governed by DSR-deformed relativistic symmetries~\cite{gacdsr1,dsrPOLAND2001},
  in two complementary regimes: longitudinal and transverse one.\\
It is valuable for our purposes to observe that ``relative locality" is a natural next step
in a logical line on relativistic theories that starts with the relativity of rest (Galilean Relativity)
and then, accommodates ``relative simultaneity" (Special Relativity).
The relativity of rest and the absoluteness of time hosted by Galilean relativity
famously leads to a
structureless law of composition of velocities: $\vec{v}\oplus\vec{\delta v}=\vec{v}+\vec{\delta v}$.
The observer-independence of the speed of light, and the relativity of simultaneity,
lead to a more complex (nonlinear, noncommutative, nonassociative)
law of composition of velocities, the special-relativistic law
$$
\vec{v}_a\oplus\vec{v}_b=\frac{\vec{v}_a+\vec{v}_b}{1+\frac{\vec{v}_a\vec{v}_b}{c^2}}+\frac{1}{c^2}\frac{\gamma_a}{1+\gamma_a}\frac{\vec{v}_a\wedge(\vec{v}_a\wedge\vec{v}_b)}{1+\frac{\vec{v}_a\vec{v}_b}{c^2}}
~,$$
where $\gamma_a=1/\sqrt{1-|\vec{v}_a|^2/c^2}$ is the Lorentz factor. In quantum-gravity research one finds motivation for endowing a specific momentum scale,
 essentially the Planck scale, with a special role, and it is conceivable~\cite{gacdsr1,dsrPOLAND2001}
 that this momentum scale, here denoted with $|\ell|^{-1}$, be an absolute relativistic scale,
 to which the laws of transformation among observers could adapt in the context of a novel
 relativistic theory of ``Doubly Special Relativity (DSR) type"~\cite{gacdsr1,dsrPOLAND2001}.

It is emerging that this sort of constructions may require~\cite{bob,kbob}
a relativity of spacetime locality in
the same sense that the observer-independence of the speed-of-light scale requires the relativity
of simultaneity. And, just like special relativity required a nonlinear law of composition of
velocities, these DSR constructions require
a non-linear law of composition of momenta $(p\oplus q)_a\neq (p+q)_a$, of the type
\begin{equation}
 p_a\oplus dq_a=p_a+U(p)^b_a dq_b=p_a+dq_a+\Gamma_a^{bc}p_b dq_c+... \label{somma}
\end{equation}
where we denoted with $\Gamma_a^{bc}$
the feature $\Gamma_a^{bc}=-\frac{\partial}{\partial p_a}\frac{\partial}{\partial q_b}(p\oplus q)_c|_{q,p=0}$, implicitly referring to the fact that this may be viewed~\cite{principle}
 as an affine connection on momentum space.\\

The liaison between our approach (which we are here going to summarize) and the one based on the study of geometry of momentum space has been investigated in \cite{jackgab}, where a first
insight of interaction vertexes in a deformed symmetries scenario is given.

We shall here not rely explicitly on the geometry of momentum space and on the form of the
composition law for momenta, since we do not need the full characterization of the shape of interactions to present longitudinal and transverse effect of relative locality. Instead we summarize
 some of the results obtained in Refs.~\cite{bob} and~\cite{transverse}, of the non-interactive case,
giving usefully complementary characterizations of the relativity of spacetime
locality that affects certain descriptions of free particles
in Planck-scale-deformed phase-space constructions.

 In this paper from now on we will formalize
the features of obstruction of measurability and modified momenta composition rule with $\kappa$-Poincar\'e Hopf algebra. The algebric sector of $\kappa$-Poincar\'e can be interpreted as a deformation
of Poincar\'e Lie algebra with $\ell\sim 1/M_P$ (where $M_P$ is the Planck mass) the parameter of deformation. It is, therefore, always possible to rely on the existence of a "classical" limit
by turning off this deformation ($\ell\rightarrow 0$). In this context moreover
 it is easier to give a simple interpretation of relative locality phenomena on physical observables and thus implement a phenomenology.

We adopt units such that the
 speed-of-light scale is $1$ ($c=1$).

\section{Longitudinal effects}

Let us start, in this section, by summarizing the main results of Ref.~\cite{bob,kbob},
where a relativistic description of momentum dependence of the speed of massless particles
was given within a classical-phase-space construction inspired by
properties of the $\kappa$-Poincar\'e Hopf algebra~\cite{lukie1992,majrue,lukieANNALS}.\\
At the quantum level the $\kappa$-Poincar\'e algebra is intimately
related to the $\kappa$-Minkowski noncommutative spacetime~\cite{majrue,lukieANNALS}. For classical-phase-space constructions indeed one can
consider standard Minkowski spacetime coordinates in combination with
a description of relativistic transformations  inspired by
properties of the $\kappa$-Poincar\'e Hopf algebra.
So  we assume trivial Poisson brackets for
the spacetime coordinates, $\{ x , t \} = 0$.
One then codifies~\cite{bob,jurekvelISOne} boost transformations
in terms of Poisson brackets with the translation generators, with deformation parameters $\alpha,\beta,\gamma$,
as follows~\cite{bob}:
\begin{eqnarray}
 &\left\{ \mathcal{N}_i, \Omega \right\} = P_i-\alpha\ell\Omega P_i\;\;\;\; , \;\;\;\;\;\;\;\; \left\{ P_i, \Omega \right\} = 0&\label{poinc1}\\
&\left\{ \mathcal{N}_i, P_j \right\} = \Omega\delta_{ij}+\ell\left((1+\gamma-\alpha)\Omega^2+\beta \vec{P}^2\right)\delta_{ij}-\ell\left(\gamma+\beta-\frac{1}{2}\right)P_iP_j~,&\label{poinc2}
\end{eqnarray}
where $\Omega$, $P_i$ and $\mathcal{N}_i$ are respectively time translation, space translation and boost generators.\\
It is easy to check that (\ref{poinc1}),(\ref{poinc2}) satisfy all Jacobi identities
with $\{ x , t \} = 0$ and standard symplectic structure.
And from (\ref{poinc1}), (\ref{poinc2}) one finds the following deformed on-shell relation:
\begin{equation}
\mathcal{C}_\ell = \Omega^2 - \vec{P}^2 +\ell( 2\gamma \Omega^3
+ (1-2 \gamma) \Omega \vec{P}^2 )~.
\label{casparam}
\end{equation}
By a standard Hamiltonian analysis one finds~\cite{bob,kbob}
from (\ref{casparam}) that, in particular,
the worldlines of  massless particles ($\mathcal{C}_\ell=0$)
are governed by
\begin{equation}
(x-x_0)_i = (1-\ell |\vec{P}|) \frac{P_i}{|\vec{P}|} \, (t-t_0)~.\label{wordli}
\end{equation}
The fact that the speed of massless particles here depends on
momentum\footnote{The coefficients of the terms $\ell \Omega^3$
and  $\ell \Omega \vec{P}^2$ in $C_\ell$ were arranged in Ref.~\cite{bob}
just so that this speed law for massless particles, $1-\ell |p|$,
would be produced. This is how from the more general two-parameter
case $\mathcal{C}_\ell = \Omega^2 - \vec{P}^2 +\ell( \gamma' \Omega^3
+ \gamma" \Omega \vec{P}^2)$
one arrives at the one-parameter case considered here and in Ref.~\cite{bob}:
$\mathcal{C}_\ell = \Omega^2 - \vec{P}^2 +\ell( 2\gamma \Omega^3
+ (1-2 \gamma) \Omega \vec{P}^2)$.}
is the main intriguing feature of this relativistic framework,
and was the subject of several investigations
(see, {\it e.g.}, Refs.~\cite{jurekvelISOne,gacMandaniciDANDREA,mignemiVEL,ghoshVEL}),
including some which established the presence of longitudinal relative
locality~\cite{bob,kbob,leeINERTIALlimit,arzkowaRelLoc}.

The relativity of locality is immediately seen
upon studying
the covariance under $\alpha, \beta, \gamma$-deformed boosts
of the worldlines (\ref{wordli}). One finds~\cite{bob} that
 an infinitesimal deformed boost with rapidity vector $\xi_i$
 acts as follows:
\begin{eqnarray}
 P_i' &=& P_i-\xi_i\Omega-\ell\xi_i(\beta P^2+(1+\gamma-\alpha)\Omega^2)-\ell\xi_k\left(\frac{1}{2}-\beta-\gamma\right)P^kP_i \label{tr1}\\
 t' &=& t-\xi_ix^i-\ell\xi_i(\alpha t P^i+2(1+\gamma-\alpha)x^i\Omega)\label{tr2}\\
 x_i' &=& x_i-\xi_i t+\ell(\alpha\xi_i t \Omega + 2\beta\xi_k x^k P_i)-\ell\left(\gamma+\beta-\frac{1}{2}\right)(\xi_kP^kx_i+\xi_ix_kP^k)\label{tr3}
\end{eqnarray}
We can now specialize  these (\ref{tr1}), (\ref{tr2}) and (\ref{tr3})
to the case of boosts along the direction $x$, which is the direction of motion
of the particle according to  (\ref{wordli}), and it then
follows that when a given observer Alice has the
particle on the worldline (\ref{wordli}) an observer boosted with respect to Alice
along the $x$ direction sees the particle on the
worldline
\begin{equation}
(x'-x'_0)_i = (1-\ell |\vec{P'}|) \frac{P'_i}{|\vec{P'}|} \, (t'-t'_0)~,\label{wordli2}
\end{equation}
indeed consistently with the relativistic nature of the framework.

And these results on the action of boosts also allow one to expose
the relativity of spacetime locality.
We first recall~\cite{bob} that the absoluteness of spacetime locality
amounts to the property  that when one observer establishes that two events coincide
 then all other observers agree that those two events coincide.\\
 To see that this is not the case in the framework under consideration,
 it suffices to consider two worldlines of massless particles with different energy, a "soft" one with momentum $p^{(s)}$ and a "hard" one with momentum $p^{(h)}$,
 emitted simultaneously from the origin of an observer Alice, so that we have a coincidence
 of (emission) events. A distant observer Bob at rest with respect to Alice would still
 describe the two emission events as coincident, but, using our prescription for boosting,
 a third observer Camilla, purely boosted with respect to Bob, would describe the two
 emission events as non-coincident.
 Details on the derivation of this relative-locality effect
 can be found in Ref.~\cite{bob}, which focused on the case of a Bob-Camilla boost parallel
 to the Alice-Bob translation direction. And for that case  Ref.~\cite{bob} exposed a ``longitudinal relative
 locality" in the sense that the nonlocality attributed to the emission events by distantly boosted
 observer Camilla was found to be along the direction connecting Camilla and
 the relevant pair of events.

\section{Transverse effects}
Our next task is to summarize the results on ``transverse relative locality" of
Ref.~\cite{transverse}.
Again we want to first map the worldline (\ref{wordli}), attributed to observer Alice,
 to the description
of an observer Bob, purely translated with respect to Alice along the $x$ direction,
but then we are interested in an observer Camilla purely boosted with respect to Bob
along a direction $y$, orthogonal to $x$.
Specializing our boost actions  to the case of a boost purely in the $y$
direction\footnote{Notice that it may be useful to also note that
 the coordinate velocity for massless particles transforms  under our boosts
 according to
$$
\{\mathcal{N}_j,(1-\ell |\vec{P}|) \frac{P_i}{|\vec{P}|} \}  =  \left( 1- \ell \left(\alpha - \beta - \gamma -\frac{1}{2}\right)|\vec{P}|\right) \left( \delta_{ij}- \frac{P_i P_j}{|\vec{P}|^2} \right) -\ell \delta_{ij} |\vec{P}|.
$$},
we arrive at this Camilla worldline, which we characterize in the $x,y$ plane~\cite{transverse}:
\begin{eqnarray}
y^C(x^C)=-\xi_y\left(1-\left(\alpha-\beta-\gamma-\frac{1}{2}\right)\ell p\right)x^C-\ell\xi_y a p
\end{eqnarray}
where $\xi_y$ is the boost parameter.
Here we notice two main features that characterize this result with respect
to the corresponding result that applies in
the special-relativity limit ($\ell \rightarrow 0$):\\
\indent (I) when $\ell\xi_y a p$ is within the reach of available experimental
sensitivities it will be appreciated
 that the worldline does not cross Camilla's spatial origin, a feature
we shall find convenient to label as ``shift";
\\
\indent (II) when $\ell\xi_y p$ is within the reach of available angular resolutions
(and $\alpha-\beta-\gamma-\frac{1}{2} \neq 0$)
the angle in the $x,y$ plane by which Camilla sees the arrival of the particle
is momentum dependent, which is a feature that may be labeled ``dual-gravity lensing",
following the terminology introduced in Ref.~\cite{leelaurentGRB}, where an analogous
feature was encountered in working with the relative-locality curved-momentum-space
framework of Ref.~\cite{principle}.\\
Both of these features in general contribute to the transverse relative locality:
if one considers a simultaneous emission at Alice of two massless particles of different
momentum our distantly transversely boosted observer Camilla ends up describing
the two emission events as non-coincident and the lack of coincidence occurs along
the $y$ direction (a direction orthogonal to the one connecting the observers Alice and Camilla).

Particularly striking are the effects of dual-gravity lensing: Alice sends out simultaneously
two particles toward Camilla along the same direction, but Camilla
ends up detecting them along non-parallel directions forming an angle~\cite{transverse}
$$
\theta \simeq \xi_y (\alpha-\beta-\gamma-\frac{1}{2}) \, \ell (p^{(h)} - p^{(s)}),
$$
where notably the angle depends linearly on the difference of the momenta $(p^{(h)} - p^{(s)})$.

Incidentally it is noteworthy that the feature
of dual-gravity lensing exposed in Ref.~\cite{leelaurentGRB} was proportional
to the sum of the energies(/momenta) of the two particles whose wordlines
were experiencing lensing. It was
already clear from Ref.~\cite{leelaurentGRB}
that this result of dependence on the sum of energies had only been checked within a
very specific setup for the derivation, including definite choices among the many possible chains of interactions that
could be considered in the interacting-particle framework there being studied. The fact that Ref.~\cite{transverse}, although within the limitations of a theory
of free particles, found dual-gravity-lensing effects proportional
to the difference of the energies(/momenta) of
the two particles whose wordlines experience lensing
can provide encouragement for the search of other chains of
interactions, in which the difference of energies governs the dual-gravity lensing.

\section{Closing remarks}
The ``theoretical evidence" we here summarized on the basis of the results originally
reported in Refs.~\cite{bob,kbob,transverse} suggests that transverse relative locality
and  longitudinal relative locality are in some ways rather different but may deserve
equal priority in future investigations of relative locality.
This intuition is based on relative-locality theories
of free particles,  but it is hard to imagine that for
interacting relative-locality particles, which are described
within the framework of Ref.~\cite{principle},
the ``balance of power" between longitudinal and transverse relative
locality could be significantly shifted.

Another aspect which was here only considered somewhat implicitly
is the relevance of relative locality for certain models with spacetime noncommutativity.
There it will also be interesting to learn how relative locality interfaces with
the most subtle aspects of those models, such as the need for adopting
consistent ordering prescriptions~\cite{aadbasis,kowanowabasis,meljaEPJ}.

Of course, we are most excited about the potential implications for phenomenology.
It was amusing to see that the possibility of DSR-deformations of relativistic symmetries
played a significant role~\cite{whataboutopera,operaDSR,synchroDSR,fransDSROPERA,yiDSROPERA,dimitriDSROPERA}
 in the analysis of the apparent ``OPERA anomaly" for neutrinos
of accelerator energies.
But the truly significant prospects are the one for Planck-scale phenomenology,
mostly exploiting observations of particles propagating over cosmological
distances (see, {\it e.g.} Refs.~\cite{magicELLISnew,unoEdue}).
The relevance of  longitudinal
relative locality for such phenomenological studies was already
established in Refs.~\cite{bob,leelaurentGRB}, and the theoretical evidence  we here summarized
may encourage attempts of seeking similar opportunities for transverse relative locality.

\end{document}